 \documentclass[twocolumn,showpacs]{revtex4}

\usepackage[latin1]{inputenc}
\usepackage{graphicx}%
\usepackage{dcolumn}
\usepackage{amsmath}

\makeatletter
\def\btt#1{\texttt{\@backslashchar#1}}%
\DeclareRobustCommand\bblash{\btt{\@backslashchar}}%
\makeatother

\begin{document}

 \title{Breakdown in vehicular traffic: driver over-acceleration, not  over-reaction }

\author{Boris S. Kerner$^1$}

 \affiliation{$^1$
Physics of Transport and Traffic, University of Duisburg-Essen,
47048 Duisburg, Germany}

\pacs{89.40.-a, 47.54.-r, 64.60.Cn, 05.65.+b}

\begin{abstract} 
Contrary to a wide-accepted  
assumption about the decisive role of driver over-reaction   for   breakdown in vehicular traffic, we have shown that  the cause of the breakdown is  driver over-acceleration, not  
  over-reaction. To reach this goal, we have introduced
	a mathematical approach for the description of driver over-acceleration in
	a microscopic  traffic flow model. The model,  in which no driver over-reaction occurs,
	explains all observed empirical nucleation features of traffic breakdown.  
 \end{abstract}

\maketitle

Traffic breakdown is a transition from free flow to congested vehicular traffic  occurring mostly at   bottlenecks.
In 1958s-1961s, Herman, Gazis, Montroll, Potts,
Rothery, and Chandler~\cite{GM_Com}
 as well as Kometani and Sasaki~\cite{KS} assumed that the cause of the breakdown
is  driver {\it over-reaction} on the deceleration of the preceding vehicle: Due to a delayed deceleration of
the   vehicle resulting from a driver reaction time
 the speed   becomes less than the speed of the preceding vehicle. If this
over-reaction  is realized for all following drivers,
then   traffic   instability occurs~\cite{GM_Com,KS,Articles}. The instability leads to a
wide moving jam (J) formation in free flow
(F) called as an F$\rightarrow$J transition~\cite{KK1994}. The traffic  instability
is currently a theoretical basic of standard traffic theory
(e.g.,~\cite{Articles,Reviews,Reviews2}).

\begin{figure}
\begin{center}
\includegraphics[width = 8 cm]{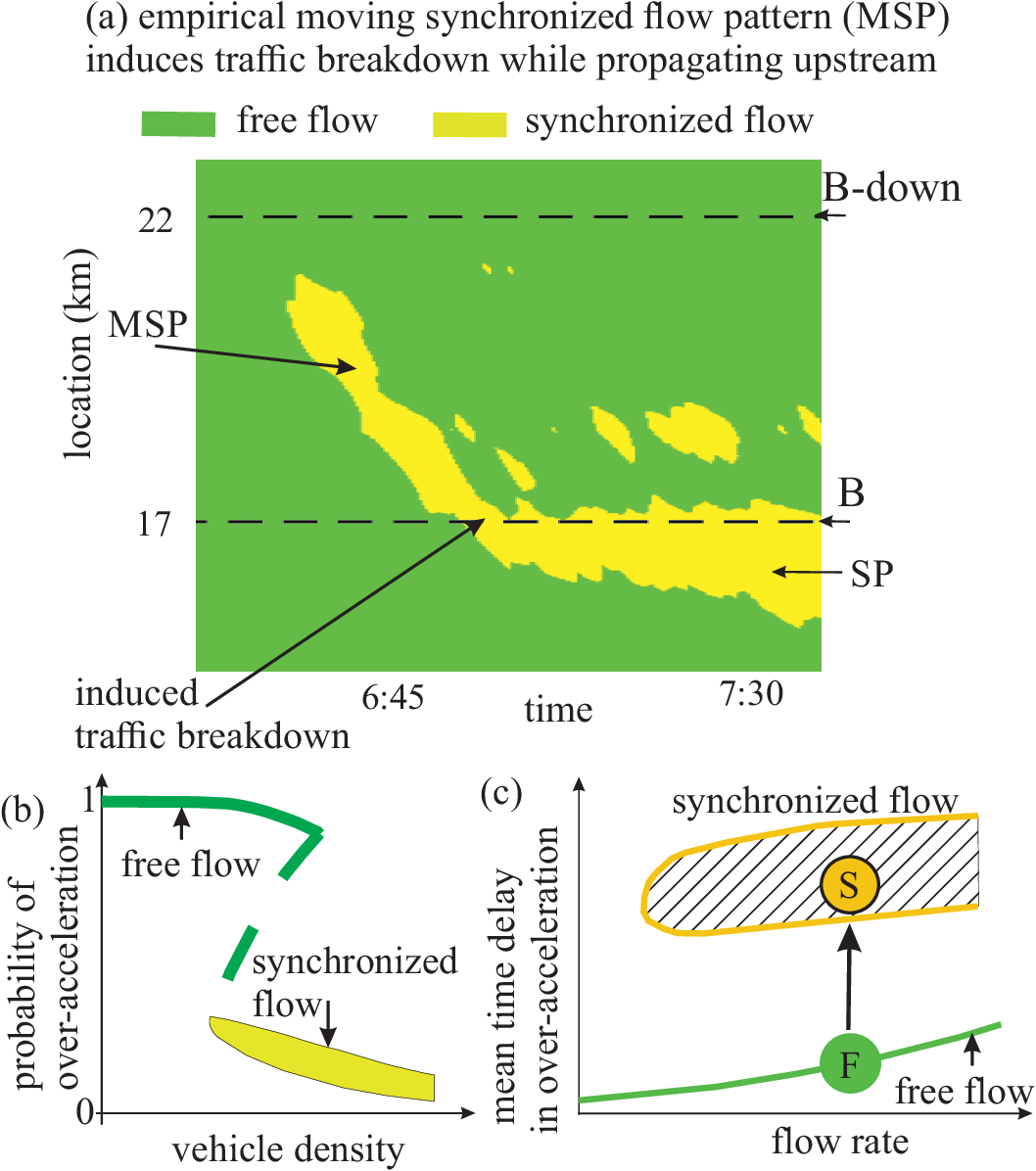}
\end{center}
\caption[]{Empirical nucleation nature of  traffic breakdown (F$\rightarrow$S transition) at bottlenecks  
(a) and hypothesis about discontinuous character of over-acceleration (b, c)~\cite{OA}. (a) Speed data    presented  
in space and time with an averaging method   were measured
with road detectors installed along   road: 
 A moving synchronized flow pattern (MSP) that has emerged   at   downstream bottleneck (B-down) while propagating upstream induces
F$\rightarrow$S transition (induced traffic breakdown) leading to
emergence of  synchronized flow pattern (SP) at upstream  bottleneck (B); adapted from~\cite{Three}.
(b, c) Qualitative density-dependence of over-acceleration
probability per a   time interval (b) and equivalent presentation of (b)
as  
 discontinuous flow-rate dependence of the mean time delay in
over-acceleration  (c);  F and S are states of free flow and
synchronized flow, respectively.
}
\label{Nucleation_Emp1}
\end{figure}

However, rather than the F$\rightarrow$J transition, in real   field  
data   traffic breakdown is a phase transition from  
free flow to   synchronized flow  (S) 
(F$\rightarrow$S transition)~\cite{Three,Three2}; the empirical traffic breakdown 
exhibits the nucleation nature (Fig.~\ref{Nucleation_Emp1}(a))~\footnote{We do not consider
classical Lighthill-Whitham-Richards (LWR) model of traffic breakdown~\cite{LWR}
because the LWR model cannot explain the empirical nucleation nature of traffic breakdown~\cite{Three}. }.
To explain the empirical nucleation nature of the F$\rightarrow$S transition, 
three-phase traffic theory was introduced~\cite{Three}, in which  
 there are  three phases: free flow (F), synchronized flow (S), and   wide
moving jam (J), where the phases S and J belong to congested traffic.

Driver over-reaction  that should explain traffic
 breakdown can occur
{\it only} if  space gaps  between vehicles are small enough~\cite{GM_Com,KS,Articles,KK1994,Reviews,Reviews2}.
 At large enough gaps, rather than over-reaction,
the   vehicle speed   does {\it not}
 become less than the speed of the decelerating preceding vehicle, i.e., usual
{\it speed adaptation}  to the speed of the preceding vehicle occurs that    causes {\it no}    instability.

\begin{itemize}
\item [--]
  Contrary to standard theory~\cite{GM_Com,KS,Articles,KK1994,Reviews,Reviews2}, it is assumed 
in three-phase traffic theory~\cite{Three} that
traffic breakdown   is realized at larger   gaps between vehicles when
 no driver over-reaction   can still occur. 
 \end{itemize}
 
In three-phase traffic theory, the empirical nucleation nature of the F$\rightarrow$S transition is explained through
a hypothesis about  a discontinuity in the probability of   vehicle acceleration when
free flow transforms into synchronized flow
 (Fig.~\ref{Nucleation_Emp1}(b))~\cite{OA}: In free flow,
 vehicles can accelerate from car-following at a lower speed to a higher speed 
with a larger probability than it occurs in synchronized flow. Vehicle acceleration that probability   exhibits
the discontinuity when free flow transforms into synchronized flow is called      {\it over-acceleration},
to distinguish over-acceleration from $\lq\lq$usual" driver acceleration  that does not  show
a discontinuous character. The    discontinuous character of over-acceleration    is explained as follows: Due to smaller space gaps  in synchronized flow, vehicles  prevent
each other to accelerate  from a local speed decrease; contrarily, due to larger space gaps     in free flow
at the same flow rate vehicles can easily accelerate from the local speed decrease.
The discontinuous character of   over-acceleration can lead to
  an S$\rightarrow$F instability in synchronized flow~\cite{Three}.
  Contrary to
	the classical traffic instability that is a growing wave of a  local {\it decrease}
in the vehicle speed~\cite{GM_Com,KS,Articles,KK1994,Reviews,Reviews2}, the S$\rightarrow$F instability  is
  a growing wave of a local {\it increase}
in the   speed~\cite{Three}.
 Microscopic three-phase   models~\cite{KKl} that   simulate
 the nucleation nature of traffic breakdown
(Fig.~\ref{Nucleation_Emp1}(a)) show also 
the classical traffic instability leading to a wide moving jam emergence. In these complex traffic models~\cite{KKl},
  both  driver over-acceleration    and   driver over-reaction
 are important.    Thus, up to now there has been no  mathematical proof that
the cause of the nucleation nature of traffic breakdown is solely over-acceleration without the influence of driver over-reaction.

 \begin{figure}
\begin{center}
\includegraphics[width = 8 cm]{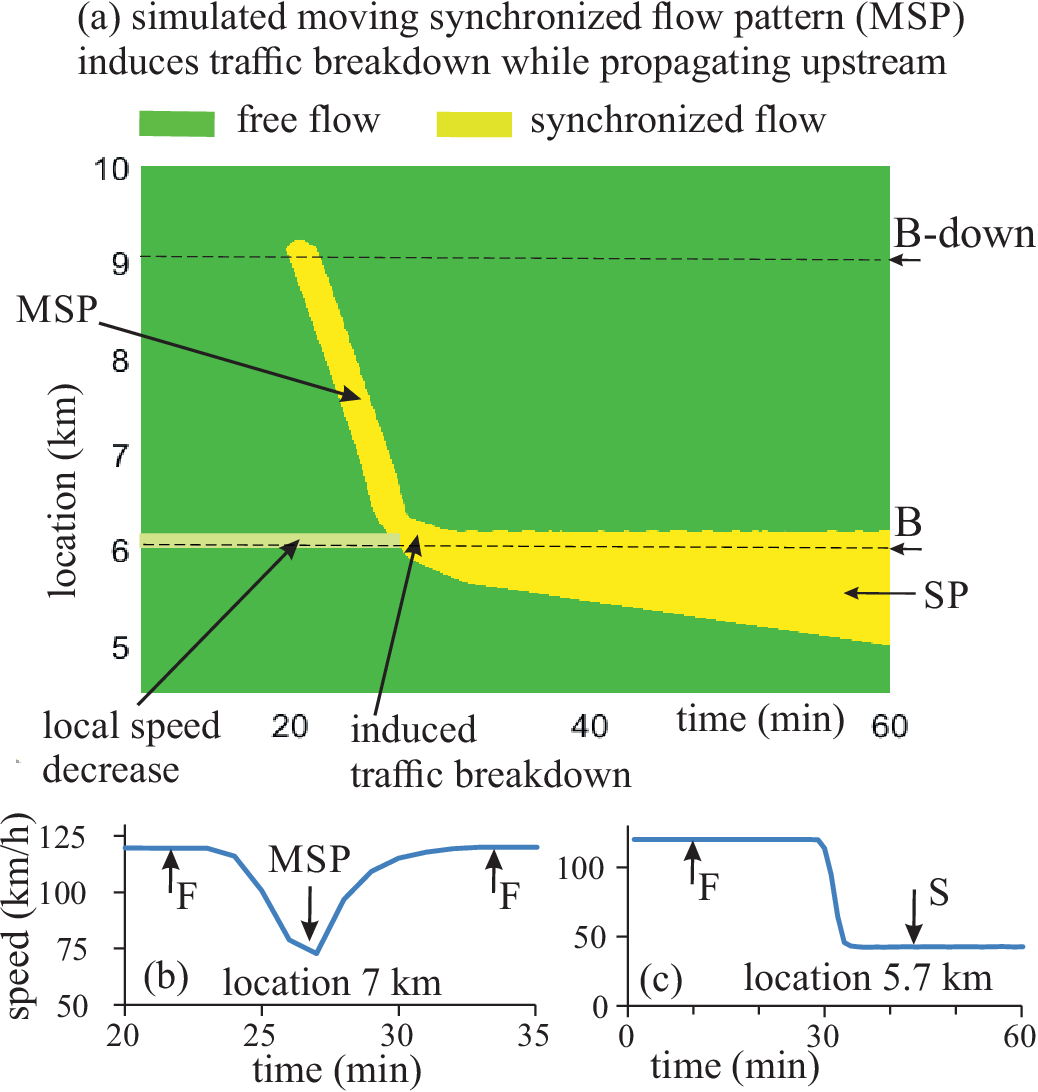}
\end{center}
\caption[]{Simulations with model (\ref{g_v_g_min1})--(\ref{a_cl}) of nucleation nature of  traffic breakdown (F$\rightarrow$S transition) on single-lane road of length $L=$ 10 km with
two  identical on-ramp bottlenecks   B and B-down at road locations $x=x_{\rm on, B}=$ 6 km
and  $x=x_{\rm on, B-down}=$ 9 km, respectively:
(a)   Speed data    presented  
in space and time as made in Fig.~\ref{Nucleation_Emp1}(a). (b, c)  Averaged (1-min) speeds 
at $x=$ 7 km within MSP (b) 
and at $x=$ 5.7 km within SP  induced through MSP propagation at bottleneck B.
Flow rate on the road at $x=0$ is 
 $q_{\rm in}=$ 2250  vehicles/h. For each of the bottlenecks that model is the same as that in~\cite{AV_2023},   there is
 a merging region  of length $L_{\rm m}=$ 0.3 km;  vehicles merge    at a middle location
  between  vehicles on the  road at  the preceding vehicle speed $v^{+}$
	when   $g>g^{\rm (min)}_{\rm target}=\lambda_{\rm b}v^{+}+ d$ with $\lambda_{\rm b}=$ 0.3 s;
on-ramp inflow rates are $q_{\rm on, B-down}=$  0 and $q_{\rm on, B}=$ 685  vehicles/h; to induce the
MSP at bottleneck B-down, impulse $q_{\rm on, B-down}=$  400   vehicles/h
at $t=$ 20 min  during 2 min has been applied. All vehicles in traffic flow are identical ones
with the following model parameters:
$\tau_{\rm safe}=$ 1 s, $\tau_{\rm G}=$ 3 s, $a_{\rm max}=$ 2.5 $\rm m/s^{2}$, $\alpha=$ 1 $\rm m/s^{2}$,
$v_{\rm syn}=$ 80 km/h, $K_{\Delta v}=$  0.8  $\rm  s^{-1}$,
$K_{1}=$ 0.15 $\rm  s^{-2}$, $K_{2}=$ 0.95 $\rm  s^{-1}$, $v_{\rm free}=$ 120 km/h, $d=$ 7.5 m. 
Under conditions $0 \leq v \leq v_{\rm free}$, vehicle motion   is found 
from    equations
$dv/dt=a$,  $dx/dt=v$  solved with the second-order Runge-Kutta method with time step 
$10^{-2}$ s.  
}
\label{Nucleation_Theory}
\end{figure}

In the paper, we introduce   a  
mathematical approach for over-acceleration   $a_{\rm OA}$:
 \begin{equation}
a_{\rm OA}=\alpha   \Theta (v - v_{\rm syn})
\label{a_OA}
\end{equation} 
 that satisfies the hypothesis
about the discontinuous character of over-acceleration (Fig.~\ref{Nucleation_Emp1}(b)).
In (\ref{a_OA}), $v$ is the vehicle speed, where $0\leq v \leq v_{\rm free}$, $v_{\rm free}$ is a maximum speed;
 $\alpha$ is a maximum over-acceleration;  $\Theta (z) =0$ at $z<0$ and $\Theta (z) =1$ at $z\geq 0$;
 $v_{\rm syn}$ is a given synchronized flow speed ($v_{\rm syn}<v_{\rm free}$). 

Based on   (\ref{a_OA}),
we develop   a microscopic traffic flow model, in which  vehicle acceleration/deceleration $a$   in a road lane
is described by
a system of equations:  
 \begin{eqnarray}
a &=&
K_{\Delta v}\Delta v + a_{\rm OA} \ \textrm{at $g_{\rm safe} \leq g \leq G$}, \label{g_v_g_min1} \\
a &=&  a_{\rm max} \ \textrm{at $ g > G$}, \label{g_v_g_min2} \\
 a &=&  a_{\rm safety}(g, v, v_{\ell}) \ \textrm{at $ g < g_{\rm safe}$}, \label{g_v_g_min3}
\end{eqnarray}
where  
$g$ is a space gap to the preceding vehicle, $\Delta v=v_{\ell}-v$,  $v_{\ell}$ is   the  preceding vehicle
speed,
$K_{\Delta v}$ is a positive   coefficient,
$a_{\rm max}$ is a maximum  acceleration, $G$ is a synchronization space-gap, $G=v\tau_{\rm G}$,
$\tau_{\rm G}$ is a synchronization time headway,
  $g_{\rm safe}$ is a safe space-gap, $g_{\rm safe}=v\tau_{\rm safe}$,
	$\tau_{\rm safe}$ is a safe time headway, $a_{\rm safety}(g, v, v_{\ell})$ is a safety deceleration. 
	The physics of model (\ref{g_v_g_min1})--(\ref{g_v_g_min3}) is as follows:

  (i)   In Eq.~(\ref{g_v_g_min1}), in addition to over-acceleration  (\ref{a_OA}), there is
	function $K_{\Delta v}\Delta v$~\cite{Three,KKl} that describes vehicle speed adaptation    to   the preceding
vehicle speed  $v_{\ell}$ occurring independent of    gap $g$ within the gap range
  $g_{\rm safe} \leq g \leq G$. Thus,
a   decrease in     $v_{\ell}$ does not lead
to a stronger decrease in  the  speed $v$: No  driver   over-reaction   occurs. 

  (ii)  Eq.~(\ref{g_v_g_min2})  describes   acceleration at large gaps $g>G$. 
	
	(iii) Contrary to over-acceleration $a_{\rm OA}$  (\ref{a_OA})  
applied	in  Eq.~(\ref{g_v_g_min1}),    function $K_{\Delta v}\Delta v$    in   Eq.~(\ref{g_v_g_min1}) at $\Delta v>0$ and Eq.~(\ref{g_v_g_min2})  describe $\lq\lq$usual"    acceleration   
 that   does not show  a
  discontinuous character.
	
	 (iv)  Eq.~(\ref{g_v_g_min3}) describes safety deceleration that should
    prevent vehicle collisions at  small gaps $g<g_{\rm safe}$; contrary to Eq.~(\ref{g_v_g_min1}), safety deceleration
		$a_{\rm safety}(g, v, v_{\ell})$ in Eq.~(\ref{g_v_g_min3}) can  lead to   driver   over-reaction. 
There are many   concepts developed in   standard models~\cite{GM_Com,KS,Articles,KK1994,Reviews,Reviews2}
		that can be used for safety deceleration
		$a_{\rm safety}(g, v, v_{\ell})$.
  For simulations below,  we
use  one of them described by Helly's function  
 \begin{eqnarray}
a_{\rm safety}(g, v, v_{\ell})=K_{1}(g-g_{\rm safe})+ K_{2}\Delta v,
\label{a_cl}
\end{eqnarray}
where $K_{1}$ and $K_{2}$ dynamic coefficients~\footnote{ Contrary to~\cite{Lee_Sch2004A},  in the 
 model   (\ref{g_v_g_min1})--(\ref{a_cl})   
  {\it  no} different states (like optimistic state or defensive state of~\cite{Lee_Sch2004A}) are assumed in collision-free traffic   dynamics governed by safety deceleration.}.

Obviously, 
through an appropriated parameter choice  
 in   standard models~\cite{GM_Com,KS,Articles,KK1994,Reviews,Reviews2} driver over-reaction is not realized
even at the smallest possible gap $g=g_{\rm safe}$ in initial steady states of traffic flow.
However, in this case no nucleation of
  congestion is possible to simulate with the standard models.

Contrarily,   if  we choose coefficients    $K_{1}$ and $K_{2}$
in  (\ref{a_cl})  (Fig.~\ref{Nucleation_Theory}) at which   even at $g\leq g_{\rm safe}$ no driver over-reaction   occurs in    model
 (\ref{g_v_g_min1})--(\ref{a_cl}), then, nevertheless, this  model shows
all known empirical nucleation features of traffic breakdown (Fig.~\ref{Nucleation_Emp1}(a)):
 An MSP induced at  downstream  bottleneck B-down
  propagates upstream. While
 reaching
upstream   bottleneck B, the MSP induces
F$\rightarrow$S transition  at the  bottleneck (Fig.~\ref{Nucleation_Theory}).
 
	\begin{figure}
\begin{center}
\includegraphics[width = 8 cm]{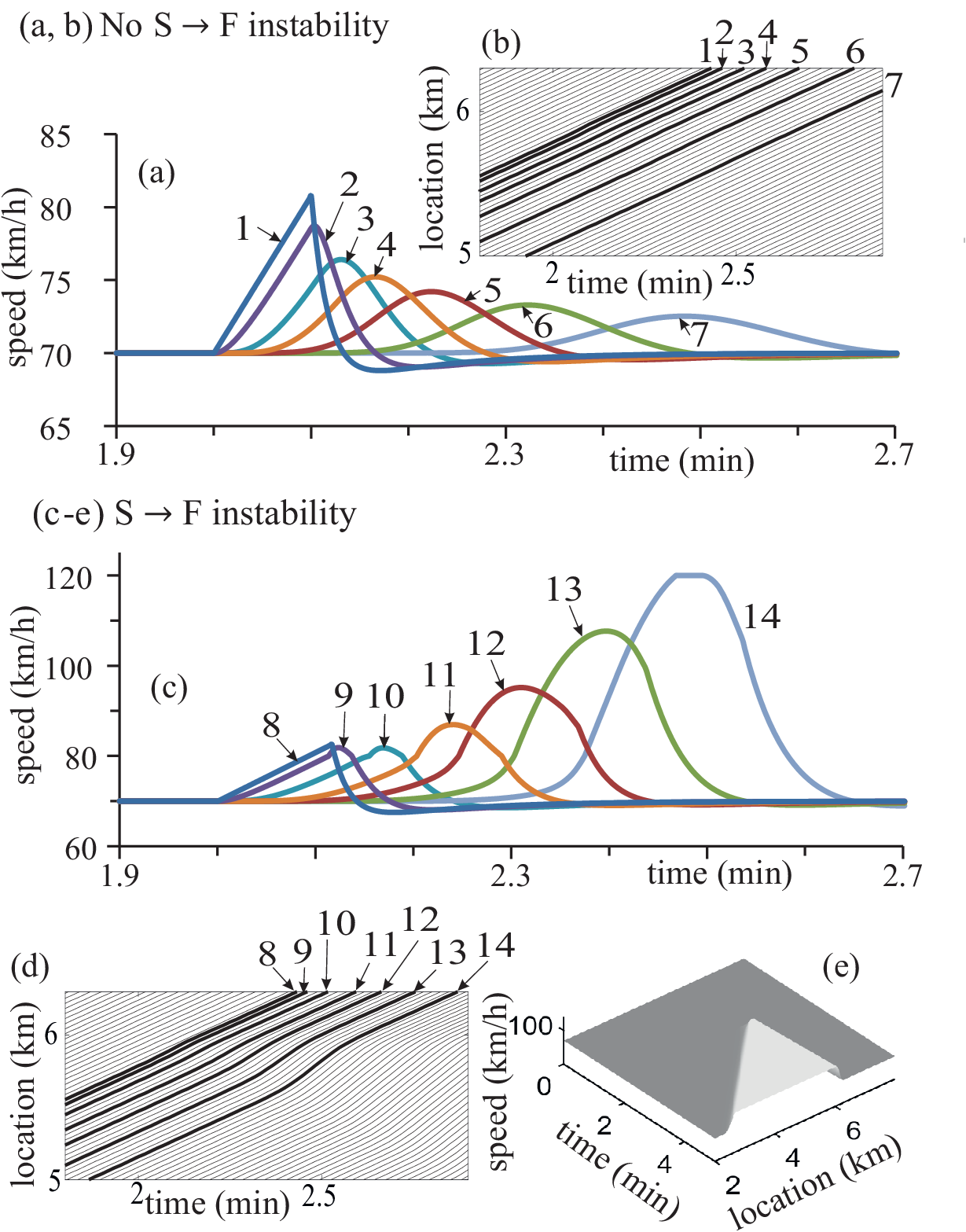}
\end{center}
\caption[]{Nucleation character of S$\rightarrow$F instability simulated
on single-lane  road (8 km long)  without bottlenecks with initial steady synchronized flow state  
 at $v=$ 70 km/h and  $g=$ 27.5 m:
(a, b) No S$\rightarrow$F instability.  (c--e) S$\rightarrow$F instability. In (a--d),
time-development of speeds (a, c) and trajectories (b, d) of  vehicles 1--7 (a, b) and 8--14 (c, d) caused by
  initial local speed increase of vehicle 1 (a, b) and vehicle 8 (c, d) simulated through vehicle
	short-time acceleration   with $a=$ 0.5
 $\rm m/s^{2}$ during   6.5 s in (a, b) and 7 s in (c, d). (e) Spatiotemporal development of speed   during
 S$\rightarrow$F instability shown in (c, d).
Other model parameters are the same as those in Fig.~\ref{Nucleation_Theory}.
}
\label{SF-instability}
\end{figure}

\begin{figure}
\begin{center}
\includegraphics[width = 8 cm]{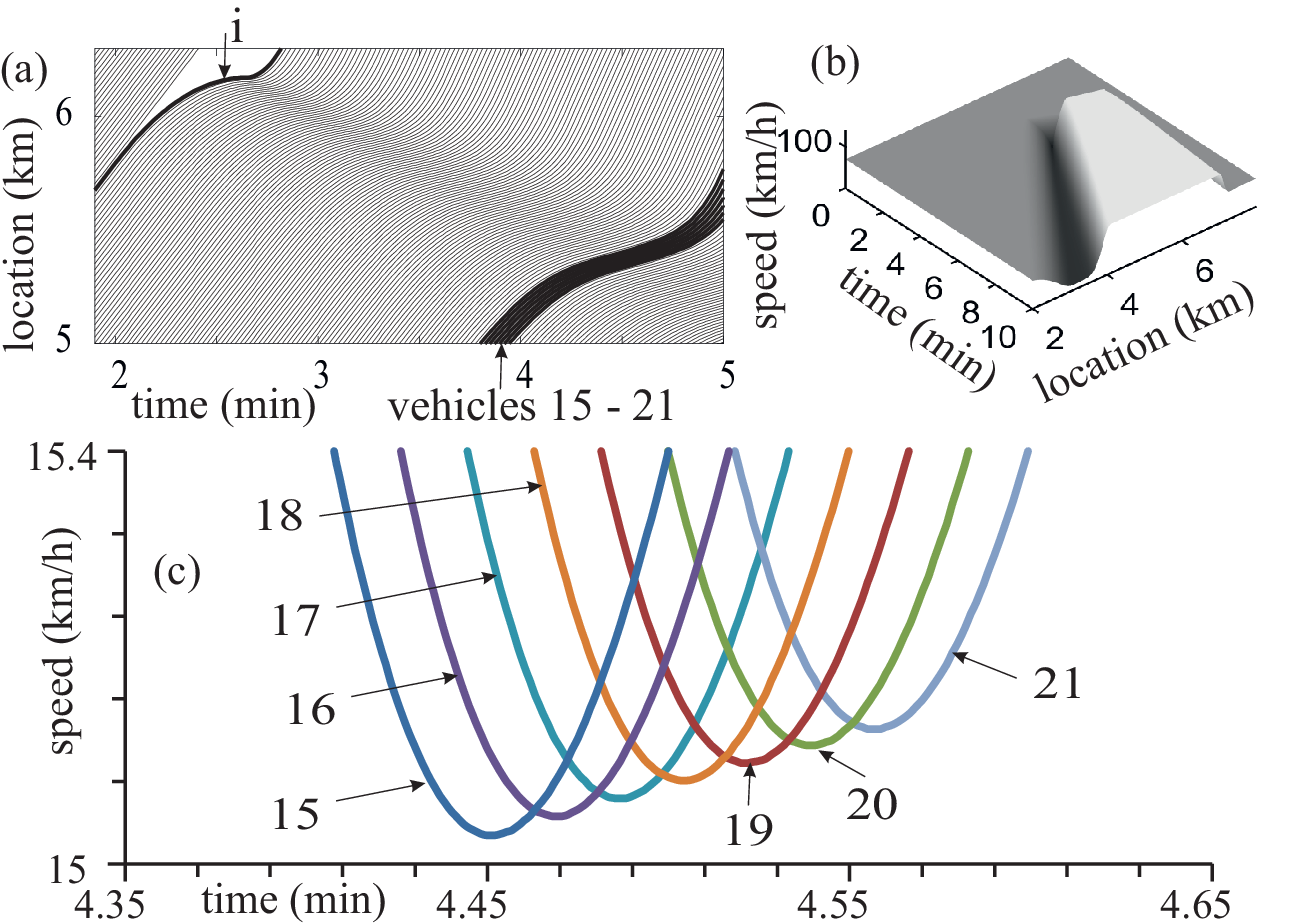}
\end{center}
\caption[]{Absence of driver over-reaction  in  model (\ref{g_v_g_min1})--(\ref{a_cl})  under
  parameters
 used in Fig.~\ref{Nucleation_Theory}.
Simulations made 
on single-lane road (8 km long) without bottlenecks with initial steady state of synchronized flow
 with $v=$ 70 km/h and $g=g_{\rm safe}=$ 19.5 m: Time-development of vehicle trajectories
(a), speed in space and time (b), and speeds of a sequence of vehicles 15--21 caused by
  initial local speed decrease of vehicle $i$   in (a)  simulated through  deceleration of vehicle $i$   with $a=-$ 0.5
 $\rm m/s^{2}$ to the speed $v= 0$; vehicle $i$ remains stationary for 1 s and then accelerates.
}
\label{No-Cl-instability}
\end{figure}

Formula (\ref{a_OA}) for over-acceleration explains induced traffic breakdown   as follows. 
Due to vehicle merging from on-ramp,    condition
 $g<g_{\rm safe}$ can be satisfied resulting in
 vehicle deceleration: A local speed decrease occurs at   bottleneck
B (Fig.~\ref{Nucleation_Theory}(a)). The minimum speed $v^{\rm (dec)}_{\rm min}$
within the local speed decrease satisfies condition $v^{\rm (dec)}_{\rm min}>v_{\rm syn}$.
Therefore, according to   (\ref{a_OA}), vehicles accelerate with over-acceleration $a_{\rm OA}=\alpha$ from
 the local speed decrease; this prevents congestion propagation
upstream of bottleneck B. Contrarily, the minimum speed within the MSP  
satisfies condition $v^{\rm (MSP)}_{\rm min}<v_{\rm syn}$ (Fig.~\ref{Nucleation_Theory}(b)). Then,
according to   (\ref{a_OA}),  over-acceleration $a_{\rm OA}=0$: When the MSP reaches bottleneck B,
synchronized flow is induced. The emergent SP remains at bottleneck B   because the speed within the SP
is less than
$v_{\rm syn}$ in (\ref{a_OA}) (Fig.~\ref{Nucleation_Theory}(c)) and, therefore,
over-acceleration  $a_{\rm OA}=0$.
 These simulations, in which
no driver over-reaction  can occur under chosen model parameters,
support the   statement of this paper:    
\begin{itemize}
\item [--] Traffic breakdown is caused by   over-acceleration, not driver over-reaction.
\end{itemize}

Formula (\ref{a_OA}) for over-acceleration explains also   the
 S$\rightarrow$F instability.   
  We consider the time-development of a local speed increase  
 in an initial steady  synchronized flow state   (Fig.~\ref{SF-instability}). The cause of the local speed increase
 is a short-time acceleration of one of  the vehicles  (vehicle  1 in
 Figs.~\ref{SF-instability}(a, b) or vehicle  8 in
 Figs.~\ref{SF-instability}(c--e)); the vehicle   must decelerate later
to   the speed of the preceding vehicle moving at  the initial   synchronized flow speed ($v=$ 70 km/h,
Fig.~\ref{SF-instability}).  
There are two possibilities: (i) The   increase in the speed of following vehicles
(vehicles 2--7 in
 Figs.~\ref{SF-instability}(a, b)) 
decays over time (Figs.~\ref{SF-instability} (a, b)); this occurs when the maximum speed
of  vehicle 2 ($v^{\rm (2)}_{\rm max}=$ 77.9 km/h) is less than $v_{\rm syn}$ in (\ref{a_OA})
and, therefore,  over-acceleration $a_{\rm OA}=0$.
(ii) Contrarily, if 
vehicle 8 (Figs.~\ref{SF-instability}(c, d)) accelerates only 0.5 s longer than vehicle 1 (Figs.~\ref{SF-instability}(a, b)), the local speed increase    initiated by
 vehicle 8
grows  over time  (vehicles 9--14 in Figs.~\ref{SF-instability}(c, d)) leading to  
  the  S$\rightarrow$F instability  
 (Figs.~\ref{SF-instability}(c--e)); this occurs because the maximum speed
of  vehicle 9  ($v^{\rm (9)}_{\rm max}=$ 81.9 km/h) is higher than $v_{\rm syn}$ in (\ref{a_OA})
and, therefore,  over-acceleration $a_{\rm OA}=\alpha$ causes the S$\rightarrow$F instability.

We have found that  in  model (\ref{g_v_g_min1})--(\ref{a_cl})  under
  parameters
 used in Fig.~\ref{Nucleation_Theory}   there is no 
driver over-reaction on the deceleration of the preceding vehicle even at the smallest possible
space gap between vehicles  $g=g_{\rm safe}$    in   an initial
 homogeneous steady state of   traffic flow. In   Fig.~\ref{No-Cl-instability},
 under condition $g=g_{\rm safe}$  in an initial synchronized flow,   vehicle   $i$ 
decelerates to a standstill, remains stationary for 1 s and then accelerates. It turns out that
none of the following vehicles decelerates to the standstill. 
The minimum speed of the  following vehicles increases slowly over time
(vehicles 15--21 in Fig.~\ref{No-Cl-instability}(c)). Finally, rather than a wide moving jam (J),
a new state of synchronized flow with  speed $v\approx$ 15.5 km/h  results
from the   deceleration of vehicle $i$.

Clearly,  other model parameters
in   (\ref{g_v_g_min1})--(\ref{a_cl})  in comparison with those
 used above (Figs.~\ref{Nucleation_Theory}--\ref{No-Cl-instability}) can be chosen at which   driver over-reaction    occurs.  
In this case, simulations of  the  model 
show   usual  results of three-phase traffic theory~\cite{Three,KKl}: (i) In free flow,  
the F$\rightarrow$S transition (traffic breakdown) occurs
that features are qualitatively the same as presented in Figs.~\ref{Nucleation_Theory}
and~\ref{SF-instability}. (ii) Contrary to Fig.~\ref{No-Cl-instability}, in synchronized flow with lower speeds 
the classical traffic instability   occurs
leading to the  S$\rightarrow$J transition. However, a detailed  analysis of these results is out of the scope of the paper.  

I   thank Sergey Klenov for help in simulations and useful suggestions.
 I  thank our partners for their support in the project $\lq\lq$LUKAS -- Lokales Umfeldmodell f{\"u}r das Kooperative, Automatisierte  Fahren in komplexen Verkehrssituationen" funded by the German Federal Ministry for Economic Affairs and Climate Action.

\end{document}